\documentclass [12pt]{article}
\usepackage{amsmath,amssymb,cite}
\usepackage{appendix}
\usepackage{feynmp}
\usepackage{float}
\usepackage[vcentermath,enableskew]{youngtab}
\usepackage{tabularx}
\usepackage[english]{babel}
\usepackage{graphicx}
\usepackage{indentfirst}
\usepackage{epsfig}
\usepackage{epstopdf}
\usepackage[]{hyperref}
\usepackage[section]{placeins}
\usepackage[stable]{footmisc}
\usepackage{appendix}
\usepackage{tabularx}
\usepackage[english]{babel}
\usepackage{graphicx}
\usepackage{indentfirst}
\usepackage{epsfig}
\usepackage{slashed}
\usepackage{fancyhdr}
\numberwithin{equation}{section}

\fancypagestyle{plain}{\fancyhead{}
}

\begin{document}

\title{
  Bell's theorem: A bridge between the measurement and the mind/body problems
}

\author{Badis Ydri\\
  Department of Physics, Annaba University, Annaba, Algeria.\\
}

\maketitle

\begin{abstract}
  In this essay a quantum-dualistic, perspectival and synchronistic interpretation of quantum mechanics is further developed in which the classical world-from-decoherence which is perceived  (decoherence) and the perceived world-in-consciousness which is classical  (collapse) are not necessarily identified. Thus, Quantum Reality or "{\it unus mundus}" is seen as both i) a physical non-perspectival causal Reality where the quantum-to-classical transition is operated by decoherence, and as ii) a quantum linear superposition of all classical psycho-physical perspectival Realities which are governed by synchronicity as well as causality (corresponding to classical first-person observes who actually populate the world). 
  This interpretation is termed the Nietzsche-Jung-Pauli interpretation and is a re-imagining of the Wigner-von Neumann interpretation which is also consistent with some reading of Bohr's quantum philosophy. 
\end{abstract}


Bell's theorem \cite{Bell:1964fg} is arguably "the most profound discovery of science" \cite{Stapp:1975} and of philosophy. This is a feeling, or for a better word an assessment, which we fully endorse in this essay. Indeed, together with the measurement problem, the collapse of the wave function, entanglement and other fundamental quantum effects, Bell's theorem lays the foundation of what we may term "experimental metaphysics" which is a genuinely possible metaphysics rooted in experimental quantum physics and not rooted in logic (which by hindsight is what Kant would have wanted). Experimental metaphysics is then a quantum metaphysics which relies on the experimental facts of quantum physics.

Bell's theorem attempt to state that the World or Reality must be logical. But physical Reality, as we know about it from scientific method, follows instead the rules of quantum mechanics, i.e.  it follows the rules of quantum logic.  Aristotelian logic states for example that if $A$ implies $B$ and if $B$ implies $C$ then $A$ must imply $C$. Bell's theorem restates this basic fact of logic as an inequality which is violated by all the experiments conducted so far \cite{Aspect:1981zz} and as a consequence this inequality must be violated by Nature itself. The propositions $A$, $B$ and $C$, in the context of Bell's theorem, are in fact sets of spin measurements performed by Alice and Bob on the electron and positron respectively which are prepared in a maximally entangled state in the EPR-Bohm-Bell experiment \cite{Einstein:1935rr,Bohm:1951xx0}. Thus, the EPR experiment and Bell's theorem are really nothing else but realistic re-formulations of the theory of quantum mechanics which nevertheless seem to contradict the nature of Nature and the real behavior of Reality.

In hindsight, Einstein in his EPR experiment wanted to undermine Bohr's interpretation of quantum mechanics (the original Copenhagen interpretation \cite{Bohr}) in favor of some form of realism (or perhaps monism), while Bohr in his well-mounted, sustained and comprehensive counter-attack  wanted to undermine quantum realism in the literal sense in favor of some form of idealism or more precisely of transcendentalism.  It is clear that Bell (with his celebrated theorem) wants (or we want him to)  sit in the middle position with an implicit (explicit) synthesis of all these elements.

The fundamental process entering the EPR experiment and Bell's theorem which is at the root of all this physical/metaphysical controversy is the quantum measurement process, i.e. the inevitable, irremovable and uncontrollable interaction between observer and observed. An almost transcendental interaction brought about by the existence of Planck's quantum of action which is the smallest possible value of action exchanged between observer and observed. As an example, the interaction between Alice and Bob from the one hand and the entangled state of the electron and positron when they arrive at the detectors from the other hand constitute a measurement event.

In fact, the measurement process involves both a psychological component (a "meaning principle") as well as a physical component (a "creation principle") as expounded in \cite{Hilgevoord-Uffink}. These components correspond to measurement-in-consciousness (conceptual terms have no meaning beyond experiments) and measurement-in-world (physical properties do not exist before observations) respectively. Hence, Reality is actually an "intentional Reality" together with a "measured Reality" which involve both the observer and the observed on an equal footing. This understanding was already highlighted by Heisenberg who believed that "the path (of a particle) comes into being only because we observe it" \cite{Heisenberg7}. This is also the picture of the measurement process obtained from Bell's theorem, e.g. the states of the system only potentially exist before the act of measurement. 

But what really constitute a measurement?

For example, do we require the observer to be classical and/or conscious separated by a Heisenberg cut \cite{Heisenberg7} from the physical process or should even the conscious observer be brought under the quantum umbrella. This is one very precise way of characterizing the measurement problem first put forward in a very dramatic way by Schrodinger and his very famous cat \cite{Schrodinger0}. But a more dramatic Gedanken experiment probing the reality/unreality of the Heisenberg cut is the one devised by Wigner and his very famous friend \cite{Wigner1}.

The quantum measurement problem is one of the most fundamental problems \footnote{If not the most fundamental problem. } in quantum physics, foundations of quantum mechanics, philosophy of physics and physics in general. The problem of the "interpretation of quantum mechanics" {\it is in fact} the "quantum measurement problem". And Bell's theorem provides, as we will discuss shortly, the framework to define rigorously the quantum measurement problem and to lay the foundation for its resolution by finding a proper interpretation of quantum mechanics. In the process we will reach another notorious problem in metaphysics and the philosophy of mind, namely the mind/body problem, and the analogy and affinity between the two problems will allow us to see novel light at the end of the tunnel in both problems. 


In summary, Bell's inequality is a very logical requirement, yet it is maximally violated by both experiments and quantum mechanics and as a consequence  it is maximally violated by both Nature and physical Reality. At this junction we can only discern two routes to proceed. In the first route we insist that physical Reality, including consciousness, is really quantum at the most fundamental level and on this view we will get the measurement problem in its strongest formulation (consciousness is part of physical Reality and physical Reality is the whole of Reality and there is no Heisenberg cut) and as a consequence we will necessarily get back to the usual interpretations of quantum mechanics and their usual problems.

In the second route we hypothesize that consciousness is a separate aspect of Reality apart from Nature but still a real aspect of Reality \footnote{A blunder of evolution as the writer of "true detectives" eloquently states.} which is identified with what we may call "psych-physical  Reality" and which should be treated classically.  On this second view, quantum mechanics is only a fundamental theory of "physical Reality" whereas classical mechanics is the fundamental theory of "psycho-physical Reality". Nature (World) is thus identified with physical Reality whereas Mind (Consciousness) is identified with psycho-physical Reality and Reality is the meshing of both physical and psycho-physical realities (\`a la Spinoza for example \cite{spinoza7}).



This second route is the one taken in this essay and we believe that Bell's theorem lends a strong support to its core thesis. We should stress that this route is a different and quite distinct one from the Wigner-von Neumann interpretation \cite{Wigner1,D47}, although they both rely on the dualism inherent in the quantum formalism (quantum dualism), with the main difference being the fact that the Wigner-von Neumann interpretation is a limiting variant of the Copenhagen interpretation  whereas the interpretation advocated in this essay is a blending of 1) the standard Copenhagen interpretation  as outlined by Dirac \cite{D47} and von Neumann  \cite{VN557} (first-person observers experience a single-world),  2) the many-world or relative state interpretation \cite{mw1,DG} (strict physicalism and causality in a many-world), 3) Bohr's original Copenhagen interpretation (Bohr's complementarity principle), 4) Nietzsche's existentialism  (perspectivism \cite{N07})  and 5) Jung's and Pauli's "unus mundus" (Quantum Reality) and synchronicity  \cite{Jung-Pauli1}). This interpretation should then be called the Nietzsche-Jung-Pauli interpretation and can be thought of as a revamping or perhaps as a re-imagining of the Wigner-von Neumann interpretation. 

Bell's theorem as we stressed above is a logical re-formulation of quantum mechanics which relies on hidden variables theories \cite{Bell:1964fg,Einstein:1935rr,Bohm:1951xx0}. Here, "logical" means that it relies on classical logic and classical mechanics which describe "psychological Reality" and {\it not} "physical Reality" and hence the dichotomy.

Bell's theorem is in fact based on the following three  very logical requirements:
\begin{enumerate}
\item {\bf Classical Realism:} This simply states that physical bodies (particles) and their properties (states) exist regardless of the acts of experimentation, observation and measurement. Thus, the world exists regardless of observation (consciousness, perception, cognition and mind).
\item {\bf Local Causality:} All causal connections must lie within the lightcone. In other words, causally-separated regions of spacetime can have no superluminal propagation of signals transmitted between them, i.e. there is no spooky action at a distance. 
  \item {\bf Free Will:} The observer can measure/observe anything she chooses without any constraints and restrictions. And the future (what the observer might measure in the future) has no bearing on the past (what the observer had actually measured). In other words, counter-factuals  (things that could happen but are contrary to what in fact actually happened) have no real effect.
  \end{enumerate}
Bell's theorem violates at least one of these logical hypotheses (which certainly or intuitively hold classically) and the eventuality that Bell's theorem violates all three of them is more than likely  \cite{Aspect:1981zz}. The measurement problem can be thought of as the problem of synthesizing coherently these three hypothesis which is not possible  by this result. The goal in the remainder of this essay is to generalize these hypotheses in way that allows a clear path to a resolution of the measurement problem and the construction of a consistent interpretation of quantum mechanics.

The first hypothesis (classical realism) is challenged/broken by the act of quantum measurement itself as summarized eloquently by Mermin when he asked (is the moon there when nobody looks?)  \cite{mermin}. Indeed, Bell's theorem together with decoherence \cite{zeh,zurek} seem to suggest that the electron (state of the electron) is not there (state does not exist) when nobody looks (before measurement) so it is natural to wonder why we can not say the same thing about the moon. The answer lies in the fact that consciousness, being the classical part of Reality,  can only perceive the classical world obtained either by the collapse of the wave function in the measurement process or via decoherence of the quantum world (this is the case of the moon who is classical via decoherence but not the case of the electron which is brought to classicality only via the act of measurement. In other words, measurement and decoherence are really two very distinct processes). Some of the most spectacular experiments in support of the breaking of the principle of (classical realism) we should mention here Wheeler's delayed choice experiment \cite{wheeler78} and the quantum Zeno effect \cite{Misra:1976by}.

The fact that (classical realism) is broken and the fact that consciousness seems to play an important part in this breaking, in the sense that we can not decouple the observed world from the conscious observer, is the first indication that there is a quantum dualism in play between the psychophysical observer and the physical observed system.  This entanglement between the observer and the observed is due to the fact that there exists in quantum mechanics a minimum value of the action  which underlies Bohr's complementarity principle \cite{Bohr}. This should be contrasted  with the situation in the theory of special relativity where the existence of a maximum value of the velocity underlies the principle of relativity.

Thus, observation in physics, according to Bohr and many other quantum physicists, is contextual and perspectival, i.e.  it is with respect and in reference to observers and in fact it requires observers.

The quantum dualism and the complementarity principle are obviously intimately related and they should be viewed as the generalization of the hypothesis of (classical realism) and the principle of causality respectively.

The second hypothesis  (local causality) is broken by the celebrated effect of quantum entanglement (or spooky action at a distance as Einstein called it).  This highly non-trivial non-local effect connects therefore regions of spacetime which are outside their mutual lightcones without violation of causality. Although, this is not a causal effect since no energy-carrying signals are propagated, this non-local effect is certainly a kind of synchronicity, between say the state of the electron as measured by Alice and the state of the positron as measured by Bob,  which carries real physical effects. Synchronicity according to Carl Jung and Wolfgang Pauli are meaningful coincidences which are not causal connections, i.e it is an acausal connecting principle (Jung) and a meaning-correspondence (Pauli).

Another synchronistic connection is the connection between the states of consciousness of the observers (Alice and Bob) and the states of matter of the particles (electron and positron). In other words, the collapse of the wave function which occurs during measurement is an acausal meaningful coincidence which is the most important manifestation of synchronicity. A remarkable experiment in support of the collapse of the wave function is  reported in \cite{PCV}  whereas quantum entanglement effects can be found in many examples some of the most intriguing are the so-called quantum games such as the CHSH quantum game (which is a particular extension of Bell's inequality) in which Alice and Bob can win $88$ per cents of the time while classically their winning can not go beyond $75$ per cents \cite{Clauser:1969ny}.

The second hypothesis (local causality) should then be generalized to a causal/synchronistic closure principle where  the causal aspects at the level of classical Reality are those of the usual local causality while at the level of physical Reality they are given by Bohr's complementarity principle. Synchronicity is the connecting principle at the level of psycho-physical Reality, i.e. those purely psychological aspects as well as some of those aspects connecting the physical and the psychological. 

The third hypothesis (free will) is known to be  challenged by causal determinism which reigns supreme in the classical world where all sorts of physical, psychological and psycho-physical arrows dominate. These arrows (action of causality, stream of consciousness, flow of time, expansion of space, diffusion of heat, propagation of radiation,  etc) are all associated with entropy increase. The most fundamental arrow which underlies all other arrows, including causal determinism, is of course the arrow of time which is identified  with the arrow of the expansion of space. The arrow of time is believed to be due to the special initial conditions of the universe found at the big bang and our current proximity to this primal event. In the quantum theory we replace determinism with indeterminism, the arrow of time with time reversal symmetry and the universe with the multiverse. A remarkable (albeit thermodynamical) model, in which the arrow of time is only understood as a local effect confined to individual universes whereas time reversal symmetry is a global feature of the larger multiverse, is the Boltzmann-Schuetz hypothesis\cite{boltzmann}.

The validity of the hypothesis of (free will) hings not so much on the elements of chance and indeterminism per se but more  on the elements of context and perspective. Thus, the existence of microscopic quantum indeterminism (of Heisenberg's uncertainty/indeterminacy principle \cite{Hilgevoord-Uffink,Heisenberg7}) does not really do anything for the existence of the macroscopic free will. Indeed, the hypothesis of (free will) is also challenged by the Free Will theorem \cite{CK} which seems to indicate that quantum indeterminism is irrelevant to free will.

However, the concept of the quantum past \cite{Hartle:1997up} in which the past (memory) and the future (free will) are unified  in the consciousness or perspective of the observer is far more promising for the existence of real free will. A far more compelling idea in this regard is the Archimedean conception of time championed by Price \cite{price} in which he makes the point that the "principle of independence of incoming influences" conflicts with time reversal symmetry in the sense that objects which will interact in the future are not independent in the same way that objects which have interacted in the past are not independent. This clearly has far reaching consequences on Bell's theorem and in particular on the hypothesis of (free will).

The act of quantum measurement is an  irreversible process associated with an increase of entropy and is therefore itself an arrow. In fact, the measurement process is a fundamental indeterministic arrow underlying consciousness and what we would like to call the "perspectival psychological time" (which is not necessarily the same as the "objective physical time"). Hence, the quantum measurement process is the arrow connecting physical and psycho-physical Realities. The intrinsic quantum indeterminism associated with the act of quantum measurement, which is quite distinct from  Heisenberg's uncertainty principle, lies also at the heart of the arrow of time and free will.

In any case, the hypothesis of (free will) must be generalized to Nietzsch's perspectivism which itself is a generalization of Bohr's complementarity principle, quantum logic and quantum contextuality. This was proposed originally by Edwards \cite{edwards7}. Thus, perspectivism in this general sense involves consciousness, free will, time, causality and synchronicity.
\begin{table}
  \begin{tabular}{|c|l|l|}
    \hline
    Bell's theorem&Bohr's quantum metaphysics&Descartes philosophy\\
    The measurement problem&Nietzsche-Jung-Pauli interpretation& The mind/body problem\\\hline
Classical realism& Quantum dualism&Dualism\\\hline
Local causality&A causal/synchronistic closure& Causal closure \\\hline
Free will& Perspectivism&Mental causation\\
\hline
  \end{tabular}
  \caption{The measurement and mind/body problems in the philosophies of quantum mechanics and mind. }\label{table}.
\end{table}

A rigorous formulation of the quantum measurement problem is then given by the three hypotheses underlying Bell's theorem (classical realism, local causality and free will) which are found to be contradicted by both experiments and quantum mechanics and hence they are inferred to be contradicted by Nature and Reality. These three hypotheses are explicitly extended to three Copenhagen hypotheses (quantum dualism, causal/synchronistic closure and perspectivism) which constitute the so-called Nietzsche-Jung-Pauli interpretation which is a re-imagining of the Wigner-von Neumann interpretation. This interpretation is believed to be also consistent with some reading of Bohr's quantum philosophy. The measurement problem is resolved within this interpretation by construction. In the table (\ref{table}) we also compare these hypotheses with the hypotheses of the mind/body problem which was articulated in its modern form by Descartes \cite{descartes7}.

\begin{figure}[htbp]
\begin{center}
\includegraphics[width=10.0cm,angle=0]{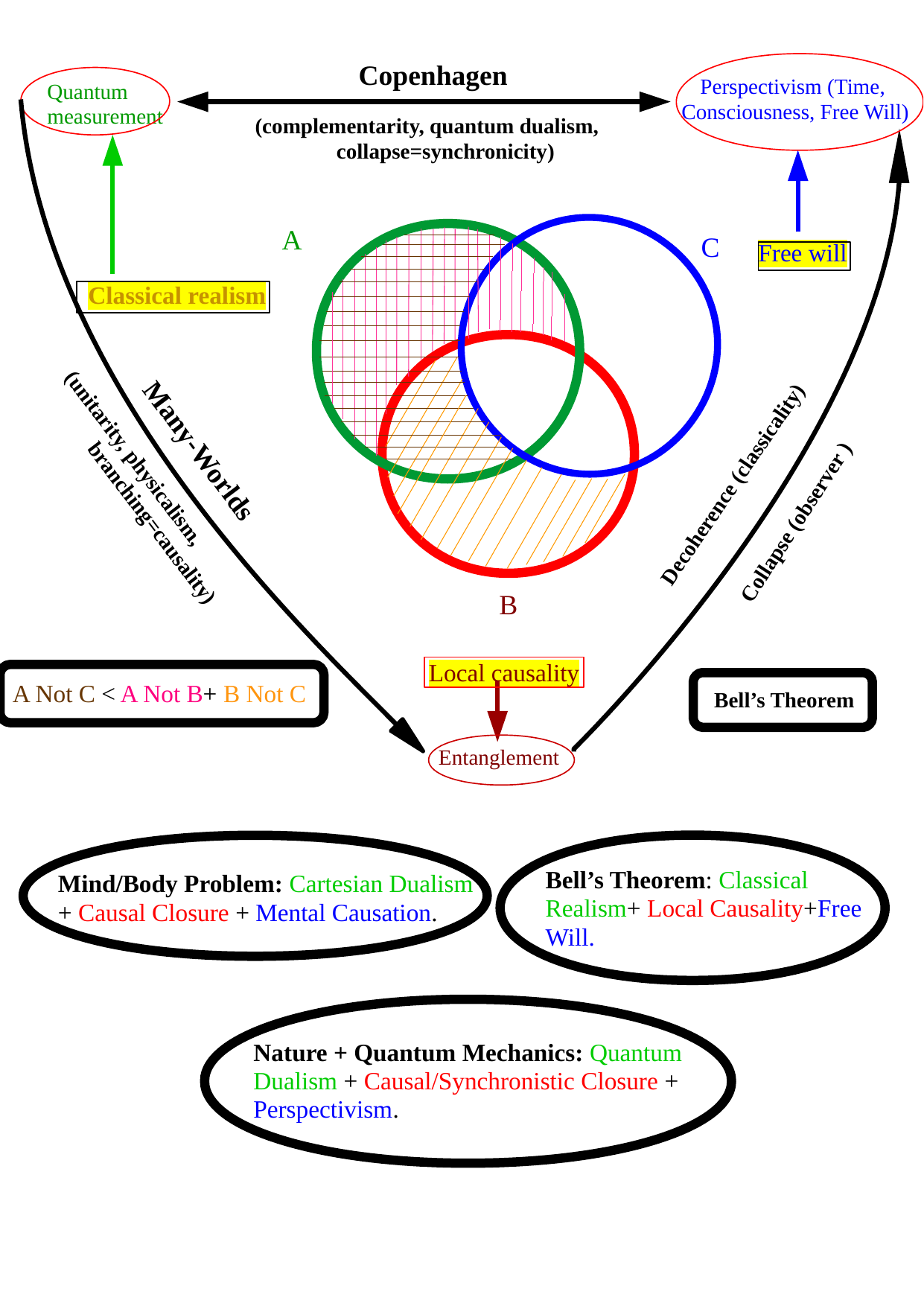}
\end{center}
\caption{Bell's theorem and the mind/body problem.}\label{bfi}
\end{figure}

In the remainder of this essay we discuss the various connecting principles relating the above three classical/logical hypotheses and their quantum/transcendental counterparts (summarized in figure (\ref{bfi})) and then we conclude with a brief discussion of the mind/body problem and how it might also be resolved within the Nietzsche-Jung-Pauli interpretation.

First, the Copenhagen interpretation with its three fundamental tenets (quantum dualism, complementarity and the collapse of the wave function) connects between perspectivism and the act of quantum measurement because this later involves an observer who is nothing but a perspective. As we have already discussed, the collapse of the wave function provides the primary example of a synchronistic connection which is a meaningful acausal coincidence between the state of the quantum physical observed system and the state of the classical conscious perspectival observer.  In fact, the collapse postulate is the basic synchronistic principle which connects in a meaningful way the coincidences between physical and psycho-physical Realities. This principle of synchronicity is a very similar principle to Leibniz parallelism \cite{leibniz7} and as a consequence the principle of quantum dualism in this context is very similar in spirit to naturalistic dualism of Chalmers \cite{chalmers7} whereas Nietzsche's perspectivism should be contrasted with Putnam's internal realism \cite{putnam}.

Second, the connection between the quantum measurement process and quantum entanglement is the many-world interpretation with its three tenets (physicalism, unitarity and superposition). Branching of the world into a multitude of worlds under successive measurements is a unitary and causal effect. Thus, there is a complementarity between the first-person observers of the Copenhagen interpretation who populate the various branches of the many-world and can only perceive their own worlds and the third-person super-observers of the many-world interpretation who can perceive the unitary causal branching and the many-world itself.   This complementarity is a generalization of Bohr's complementarity, which relates only the first-person observers of the Copenhagen interpretation, which was proposed in \cite{Susskind:2016jjb11}.

Third, the connections between quantum entanglement and perspectivism are decoherence and the collapse of the wave function. Thus, decoherence and the world-from-decoherence (classical world) from the one hand and the collapse of the wave function and the world-in-consciousness (observed world) are really two not necessarily the same classical descriptions of the objective/perceived Reality.

Quantum dualism is thus a very compound concept involving the duality between the physical and the psycho-physical, the duality between the perspectival/objective nature and the synchronistic/causal association/action of consciousness in an individual first-person observer, the complementarity between the various perspectives found in the various contexts corresponding to a single first-person observer, the complementarity between the different perspectival and synchronistic first-person observers of the Copenhagen interpretation and the complementarity between these first-person observers who populate the World and the third-person super-observers of the many-world observers who can perceive quantum Reality as a linear superposition.

Here we reach the end of our discussion of Bell's theorem, the quantum measurement problem and the Nietzsche-Jung-Pauli interpretation. The basic logical hypotheses (which are inspired by classical logic, classical mechanics and perceived by classical consciousness) of Bell's theorem are given by the three tenets (classical realism, local causality and free will).  The quantum measurement problem can be seen as the conflict between these three very reasonable requirements. As we have discussed, Nature and quantum mechanics both of them violate Bell's theorem. In other words, physical Reality as opposed to psycho-physical Reality is not subjected to classical logic but to quantum logic.

  We can immediately observe here a stark resemblance between the above hypotheses of Bell's theorem (or the hypotheses underlying the quantum measurement problem) and the hypotheses underlying the mind/body problem which are given by the three tenets: 
  \begin{enumerate}
  \item {\bf Dualism:} The only two substances constituting Reality are matter (space) and consciousness (thought). This is the analogue of the hypothesis of (classical realism). 
  \item {\bf Causal closure:} There exists only causal links between events. This is the analogue of the hypothesis of (local causality).
    \item {\bf Mental causation:} There exists a genuine effect of consciousness/mind on matter. This is the analogue of the hypothesis of (free will).
    \end{enumerate}
  It seems that Nature, in the sense of physical Reality, violates also these hypotheses which implies that not everything is physical but Reality is also psycho-physical. In fact, quantum mechanics seems to suggest in place of the three above classical hypotheses, in both the measurement problem and the mind/body problem, the alternative quantum hypotheses (quantum dualism, causal/synchronistic closure and perspectivism). These quantum hypotheses resolves the quantum measurement problem by construction so it is very possible and very plausible that they might also lead to a resolution of the mind/body problem.

A detailed account of many of the ideas outlined in this essay can be found in a greater detail in the recently published book \cite{Ydri3}. The essays \cite{Ydri:2020zjg,Ydri:2020nys,Ydri:2018ork} contain further discussion of many other related ideas found also in the book.

\end{document}